\documentclass[superscriptaddress, aps, prx, reprint, floatfix]{revtex4-2}
\usepackage{graphicx}
\usepackage{bm}
\usepackage{xspace}
\usepackage{textcomp}
\usepackage{siunitx}
\usepackage{amsmath}
\usepackage{amssymb}
\usepackage{gensymb}
\usepackage{xcolor}
\usepackage{enumerate}
\usepackage{mathtools}
\usepackage{booktabs}

\renewcommand{\eqref}[1]{Eq.~(\ref{#1})}
\newcommand{\Meth}{Materials and methods}
\newcommand{\App}[1]{Appendix #1}
\newcommand{\para}{{\mkern3mu\vphantom{\perp}\vrule depth 0pt\mkern2mu\vrule depth 0pt\mkern3mu}}

\newcommand{\umm}{\textmu{m/min}\xspace}
\newcommand{\um}{\textmu{m}\xspace}
\begin{document}

\title{Coherent Turning Behaviors Revealed Across Adherent Cells}
\author{Yiyu Zhang}
\thanks{These authors contributed equally to this work.}
\affiliation{Beijing National Laboratory for Condensed Matter Physics, Institute of Physics, Chinese Academy of Sciences, Beijing 100190, China}
\affiliation{School of Physical Sciences, University of Chinese Academy of Sciences, Beijing 100049, China}

\author{Xiaoyu Yu}
\thanks{These authors contributed equally to this work.}
\affiliation{Beijing National Laboratory for Condensed Matter Physics, Institute of Physics, Chinese Academy of Sciences, Beijing 100190, China}
\affiliation{School of Mechanical Engineering \& Automation, Beihang University, Beijing 100191, China}

\author{Boyuan Zheng}
\affiliation{Beijing National Laboratory for Condensed Matter Physics, Institute of Physics, Chinese Academy of Sciences, Beijing 100190, China}
\affiliation{School of Physical Sciences, University of Chinese Academy of Sciences, Beijing 100049, China}

\author{Ye Xu}
\affiliation{School of Mechanical Engineering \& Automation, Beihang University, Beijing 100191, China}

\author{Qihui Fan}
\email{fanqh@iphy.ac.cn}
\affiliation{Beijing National Laboratory for Condensed Matter Physics, Institute of Physics, Chinese Academy of Sciences, Beijing 100190, China}

\author{Fangfu Ye}
\email{fye@iphy.ac.cn}
\affiliation{Beijing National Laboratory for Condensed Matter Physics, Institute of Physics, Chinese Academy of Sciences, Beijing 100190, China}
\affiliation{School of Physical Sciences, University of Chinese Academy of Sciences, Beijing 100049, China}
\affiliation{Wenzhou Institute, University of Chinese Academy of Sciences, Wenzhou, Zhejiang 325000, China}
\affiliation{Oujiang Laboratory (Zhejiang Lab for Regenerative Medicine, Vision and Brain Health), Wenzhou, Zhejiang 325000, China}

\author{Da Wei}
\email{weida@iphy.ac.cn}
\affiliation{Beijing National Laboratory for Condensed Matter Physics, Institute of Physics, Chinese Academy of Sciences, Beijing 100190, China}

\begin{abstract}
Adherent cells have long been known to display two modes during migration: a faster mode that is persistent in direction and a slower one where they turn. Compared to the persistent mode, the turns are less studied. Here we develop a simple yet effective protocol to isolate the turns quantitatively. With the protocol, we study different adherent cells in different morphological states and find that, during turns, the cells behave as rotors with constant turning rates but random turning directions. To perform tactic motion, the cells bias the sign of turning towards the stimuli. Our results clarify the bimodal kinematics of adherent cell migration. Compared to the rotational-diffusion-based turning dynamics - which has been widely implemented, our data reveal a distinct picture, where turns are governed by a deterministic angular velocity.
\end{abstract}
\maketitle

\section{Introduction}

Cell migration is the basis of various physiological processes~\cite{Trepat2012}. When migrating, cells have long been found to alternate between a persistent and a non-persistent mode~\cite{Gail1970, Hall1977}. The persistent fractions are sometimes referred to as 'runs', which are driven by directional flow of actin filaments inside the cells~\cite{Murrell2015}. The actin flow serves a pivotal role when a cell runs: it breaks the directional symmetry as the cell tends to migrate along the flow's direction; meanwhile, the flow's speed governs how fast the cell can move~\cite{Maiuri2015}. Moreover, the directional flow of actin and cell polarization enhance one another. Altogether, these interactions between cell speed, polarity, and actin flow result in a universal coupling between the cells' speed and persistence (UCSP)~\cite{Maiuri2015}. This model has been a powerful tool, as it offers a comprehensive framework that encompasses both sub-cellular and single-cellular dynamics. 

Compared to the runs, the less persistent periods of cell migration, which are primarily responsible for the turning, are less studied~\cite{Allen2020,Zhang2023microglia,Jiang2023}. However, they are no less important for cell migration. A primary reason for the scarcity of studies is that kinematics of turns are more complex than that of runs. While runs appear unambiguously as straight motions and this applies universally for different microorganisms (not limited to adherent tissue cells), turning behaviors manifest in a wide variety of forms. This is best exemplified by the colorful terms coined to describe microbial locomotion, such as 'run-and-tumble'~\cite{Berg1972,Zhang2023microglia}, 'run-and-circle'~\cite{Abaurrea2017,Allen2020}, and 'run-reverse-flick'~\cite{Xie2011}. Even within eukaryotic cells, turning may mean either a period of stronger rotational diffusion~\cite{Alessandro2017,Zhang2023microglia} or a gradual but steady change in the cell's direction~\cite{Jiang2023,Allen2020}. Consequently, in experiments using different cells, turning behaviors are characterized in different ways, such as fitting to diffusive models~\cite{Alessandro2017,Shaebani2020} or frame-to-frame comparison of directions~\cite{Werner2019,Begemann2019}. All the aforementioned varieties add up and make cross-comparison of results over cell types challenging, hindering us from obtaining a holistic view of how cell turns.

In this study, we focus on adherent eukaryotic cells and we aim to resolve their common turning kinematics. A unified framework is developed to assess turning behaviors across these cells. We find that, despite variability in cell type, size, and morphology, they all turn analogously to rotors with randomized direction (chirality) but constant turning rates. Experimentally, the following picture of cell migration is revealed. First of all, the cells switch between runs and turns at constant probability rates (i.e., 2-state Markovian process featured by exponential state interval distributions). While running, the angular diffusion is strongly suppressed so that the cell can move persistently along a specific direction. Upon the finish of a run, the sign of the following turn is randomly determined and does not change thereafter. Then, during the entire turn, the cell's direction of motion changes at a constant rate. Finally, when the turn ends, the ending direction will be taken by the run that follows.  
In this picture of migration, we further resolved that the cells achieve targeted motion (mechanotaxis in this case) by biasing the probability of turning signs.
These observed kinematics offers a renewed and generalized perspective to model cell migration, with which we examine the UCSP model with an unequivocal definition of persistence. Moreover, a positive dependence between the turning rate and duration is also found over different cells, supporting the existence of actin-involved positive feedback loops hypothesized previously~\cite{Allen2020}. Our findings highlight the general applicability of constant-rate turns in adherent cell migration and contribute to a refined description of adherent cell kinematics.


\section{Materials and methods} 

\subsection{Cell culture}
Non-tumorigenic epithelial cells (MCF-10A) cells are cultured in DMEM/F12 (Sigma-Aldrich and Invitrogen) at 37\textcelsius\ in a 5\% CO$_2$ atmosphere. Supplements added to the media include 5\% horse serum (Kang Yuan Biology and Invitrogen), Pen/Strep (100$\times$ solution, Gibco, 1\% v/v), EGF (20 ng/mL, Peprotech), Hydrocortisone (0.5 \textmu{g/mL}, Sigma-Aldrich), Cholera toxin (100 ng/mL, Sigma-Aldrich), and Insulin (10 \textmu{g/mL}, Sigma-Aldrich). 

Fibroblasts (NIH-3T3) are cultured in DMEM (Corning) supplemented with 10\% bovine calf serum (GIBCO) and 1\% penicillin-streptomycin (100$\times$ solution, Invitrogen) at 37\textcelsius\ in a 5\% CO$_2$ atmosphere.

The MCF-10A cells examined on collagen substrates are kindly provided by Yang Gen Lab, Peking University, China; while the MCF-10A and NIH-3T3 cells examined on PDMS substrates are obtained from the Chinese National Biomedical Cell-Line Resource. The latter group of MCF-10A cells express a green fluorescent protein (GFP) and the NIH-3T3 cells are labeled with GFP through transduction of HBLV-ZsGreen-PURO (Han-bio).
  
\subsection{Preparation of PDMS substrates}
The silicone elastomer PDMS (Sylgard 184, Dow Corning) is cast on a dish and cured at 60~\textcelsius\xspace for 4 h (base stiffness $\sim$750 kPa). The substrate is rinsed with ethanol and blown dry with nitrogen gas. The substrate surface is further oxidized by air plasma clean (Harrick Plasma) and coated with fibronectin (50 \textmu{g/mL}, Sigma) in DPBS to enhance cell adhesion.

\subsection{Motility assays}
200 \textmu{l} MCF-10A cell suspension ($1\times10^4$ mL$^{-1}$) is deposited onto a 2 mg/mL collagen gel substrate in a 14 mm well, incubated overnight, and observed thereafter. Images are taken every 2 min for 6 hour. We focus on single cell migration and thus stop tracking when the cell attaches to another or divides into two. We further exclude the tracks that exhibit low motility (i.e. the maximum displacement during the entire recording is less than 45 \um, $\sim$1.5 cell size). Finally, $N$=396 (out of 518) MCF-10A tracks on collagen are collected. Tracking is based on template matching with customized software [OpenCV-python package (ver.4.5.1)]. 

MCF-10A and NIH-3T3 observed on PDMS substrates are observed and tracked with fluorescence microscopy. Cell suspensions of $1\times10^5$ mL$^{-1}$ concentration are used and observation starts 6 h later. Images are taken every 4\,min for 12 hours. The software Imaris is used to track the time-dependent positions of the centroids of individual cells.

All cells are observed in their maintaining condition with an inverted microscope (Nikon, Eclipse Ti).

\subsection{Mechanotaxis assays}
Needles held by a motorized micromanipulator (Eppendorf, InjectMan 4) and inserted into the collagen gel substrate to perform cyclic stretching. The needle is placed 60\,\um to the right of the targeted cell ($+x$-axis). It enters 50\,\um deep into the substrate, and pulls the substrate 30\,\um away from the cell. A stretch cycle consists of the following phases: pull (10 \textmu{m/s}), hold (30 s, 60 s, or $\infty$), release(50 \textmu{m/s}), and return (to the initial position, 100 \textmu{m/s}). The collagen gel used is uniformly mixed with 0.013\% w/v microbeads (0.87\,\um diameter, Spherotech, FP-0856-2), which help track the gel deformation. In the mechanotaxis assays, MCF-10A cells are tracked manually and the gel deformation is subtracted from the track. 

Cells are observed following the same protocol for motility assays. MCF-10A cell suspension of a lower density ($5\times10^3$ mL$^{-1}$) is employed.
Prior to the application of mechanical stimuli, the cells' motility are monitored for 20 minutes. Approximately half of the cells are found to be amoebic and in the ‘run’ state. Among the motile ones, a cell moving at an angle $\phi$ between $ \frac{\pi}{6}$ and $ \frac{\pi}{2}$ with respect to the +$x$-axis is randomly chosen for mechanotaxis assay.
Cells under different cyclic stretching (N$\approx$30 for each group) are qualitatively the same and are therefore combined. $N$=93 cells are eventually recruited.

\begin{figure*}[htbp]
    \includegraphics[width=0.97\textwidth]{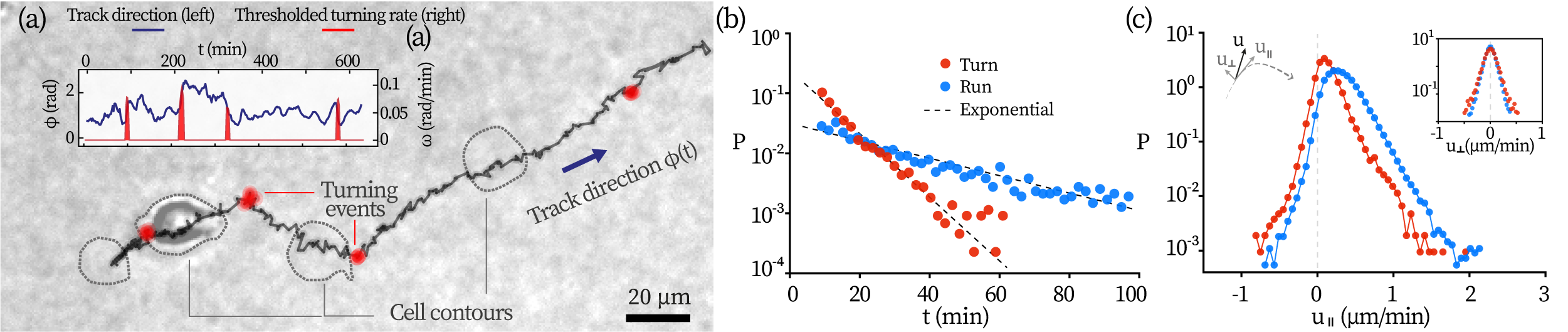}
    \caption{Kinematics of MCF-10A migraWtion on collagen substrates. (a) A typical track is displayed with the 'turns' marked red. The track's direction of motion and its thresholded turning rates are displayed in the inset. (b) The distribution of the state intervals. (c) The distributions of the cells' instantaneous velocities. The schematic drawing to the left displays how the velocity components are defined.} \label{fig:twoState}
  \vspace{-3mm}
\end{figure*}

\subsection{Methods}
The ACFs of turns presented in Figs.~\ref{fig:rotor}(e), \ref{fig:mesenchymal}(d), and \ref{fig:mesenchymal}(j) are obtained by averaging over multiple single-turn ACFs. As the turns last for different lengths of time, each scatter point on the graph may represent an average value derived from samples of varying sizes. The data is truncated at the time interval $\Delta t$ if the number of available turns are less than $N=10$ [Figs.~\ref{fig:rotor}(e)] or less than $N=5$ [Figs.~\ref{fig:mesenchymal}(d) and \ref{fig:mesenchymal}(j)].

The aspect ratio, AR, of a given cell is reported as follows. First, with image analysis we obtain the cell's contour for each frame of video. For each frame, we compute the aspect ratio of the smallest bounding box of the cell's contour, $\rm{AR}(t)$. Lastly, AR is reported as the mean of $\rm{AR}(t)$ over the entire track duration.

In mechanotaxis assays, the temporal distribution of turns of arrived cells in Fig.~\ref{fig:taxis}(e) is adjusted to account for the varying sample size (number of tracks) over time. The number of available tracks decreases over time as more and more cells have arrived at the pipette and their tracks have ended. To account for this, we present the adjusted temporal distribution of turns $\rm{PDF}^* = \rm{PDF}/\gamma(t)$, where the coefficient $\gamma(t) \equiv N(t)/N(0)$ and $N(t)$ denotes the number of still available tracks at time $t$. Moreover, the data is truncated at $t\approx 100$ min because there are fewer than $N=10$ remaining tracks thereafter. 

In studying the correlation between event duration and the mean (angular) speed (Figs.~\ref{fig:positiveFeedback} and \ref{fig:positiveFeedback_omega}), data are binned by event duration ($t_{\rm run}$ or $t_{\rm turn}$) and the latter is presented as the $x$-axis. We avoid binning by the mean speed during an event. This is because extremely large (angular) speeds induced by transient fluctuations (e.g., morphological changes), are more likely to dominate short events. Therefore, instead of averaging out the effect of noise, binning by speed actually highlights the noise, especially at large speeds.

\section{Results}
\subsection{Distinguishing the two states by turning rates}

We begin with examining epithelial cells (MCF-10A) on collagen (2 mg/mL) substrates. After the cells have adhered to the substrate, image sequences are taken every 2 min for 6~h. From the video recording, we track the cell contours for each frame and report the geometric center as the cell's location. In total, $N$=396 motile tracks of single cells are collected, for details see \Meth.

A typical track together with detected cell contours are displayed in Fig.~\ref{fig:twoState}(a) and Supplementary Video (SV).~1 (see more tracks in Fig.~\ref{fig:demo_track}). Similar to previous reports~\cite{Potdar2009}, cells migrate in a pattern where persistent motions are interrupted by brief turning events. At first sight, this pattern resembles the 'run-and-tumble' motion displayed by bacteria~\cite{Berg1972} and immune cells~\cite{Zhang2023microglia}. We have therefore attempted to discern turns from runs by speed-based criteria, which are
previously applied to determine tumble events~\cite{Berg1972,Alon1998,Zhang2023microglia}. The attempt was not successful because: (1) the cells' moving speed while turning is only slightly lower than the speed of their runs. (2) The cells' varying shapes (e.g. extension and retraction of protrusions) give rise to a transient high-speed component during both turning and running, which misleads the speed-based state marking algorithms.
   
However, we find that turning-rate-based algorithms effectively separate the two states. We compute the track direction $\phi(t)$ and its rate of change $\omega=d\phi/dt$ from the smoothed tracks, see the inset of Fig.~\ref{fig:twoState}(a). Turning events are marked by thresholding the absolute turning rate $\left|\omega\right|>\Omega_c$. This simple algorithm, requiring minimal tuning of 
$\Omega_c$, works successfully across various cell types, ranging from tissue cells (e.g., MCF-10A and NIH-3T3) to cancer cells (e.g., MDA-MB-231). A generalized protocol for determining $\Omega_c$ and more information about the algorithm is detailed in \App{A}.

We find that the duration of runs and turns both follow exponential distributions, Fig.~\ref{fig:twoState}(b). The characteristic times for runs and turns are $\tau_{\rm turn}=8.2$ min and $\tau_{\rm run}=29.9$ min\footnote{Runs longer than $\sim$150 min seem to follow either another exponential distribution with larger characteristic time ($\sim$80 min) or a long tail (e.g., power-law) distribution}. In other words, cells spend approximately 80\% [$\tau_{\rm run}/(\tau_{\rm turn}+\tau_{\rm run})$] of the time in persistent motion. This is starkly different from the motion of non-adherent cells on the same substrate, which appear to spend 80\% of the time in a stationary tumbling state~\cite{Zhang2023microglia}. 

Figure~\ref{fig:twoState}(c) presents the speed distributions of the MCF-10A cells. To better resolve the kinematics, we further decompose the cells' instantaneous velocity $\bm{u}$ along its direction of motion ($u_{\para}$) and perpendicular to the direction ($u_{\perp}$). While the instantaneous velocities are computed from frame-to-frame displacement, the direction of motion is computed from the cell's net migration over $N=7$ frames, which corresponds to a sampling window of $t_{\rm win}=12$~min. This helps isolate the cell's net migration from the displacement resulting from its fast deformations. Details about the use of $t_{\rm win}$ can be found in \App{A}.
Directions are illustrated by the schematic drawing to the left of Fig.~\ref{fig:twoState}(c). The parallel component, $u_{\para}$, generates effective forward migration. While $u_{\para}$ is indeed larger for runs, its distributions for runs and turns largely overlap. We also see that there is a finite probability for $u_{\para}$ to be negative and this marks the transient components due to the contraction of the cells' contours. The probability of having negative $u_{\para}$ is higher when a cell turns, suggesting that more significant morphological changes are taking place during then. This is further supported by a stronger fluctuation (standard deviation) in contour perimeter during turns. For both runs and turns, the perpendicular component $u_{\perp}$ distributes symmetrically with respect to 0, see the inset of Fig.~\ref{fig:twoState}(c). $u_{\perp}$ is larger during turns (broader distribution) because it contributes to changes in the direction of migration.

\subsection{Cells turn with random signs but at a constant rate}
\begin{figure}[htbp]
\includegraphics[width=0.48\textwidth]{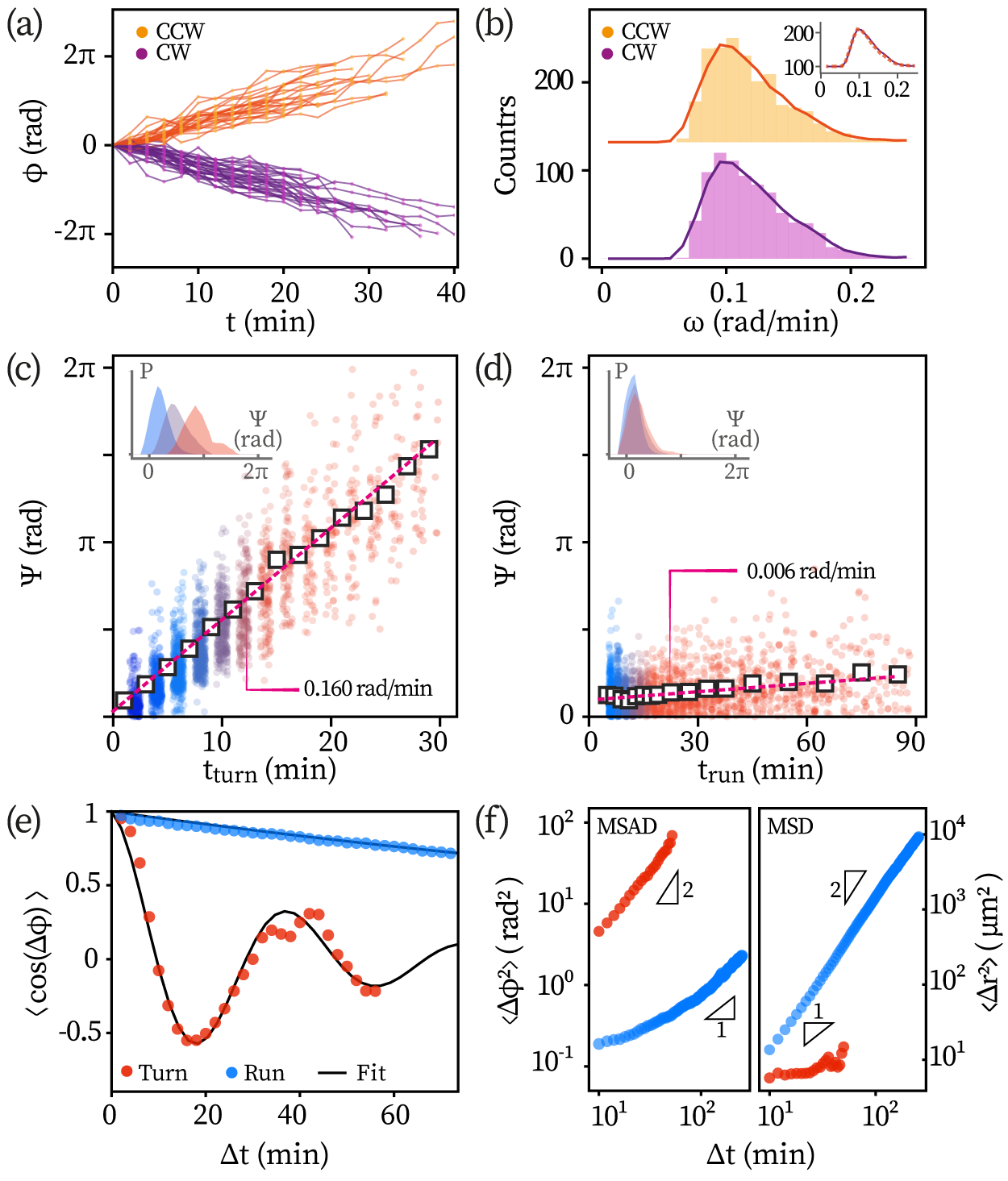}
    \caption{Kinematics of turning.
    (a) How the migration directions $\phi$ vary with time during turns. CCW: counter-clockwise, CW: clockwise. (b) Probability distribution of turning rates. Inset: the histograms' contour overlaid. The total angular change during an event $\Psi$ as a function of the event's duration ($t_{\rm turn, run}$), with (c) for turns and (d) for runs, respectively. N$\sim$2000 runs and turns are shown. Inset: probability distribution of $\Psi$ for short, intermediate, and long events. (e) The angular correlation functions evolving over time. (f) Mean squared angular displacement (MSAD, left) and mean squared displacement (MSD, right) during the two states.} \label{fig:rotor}
  \vspace{-3mm}
\end{figure}

How do cells turn? From a kinematic perspective, the answer appears to be surprisingly simple: they turn at a constant rate. Figure~\ref{fig:rotor}(a) presents how the direction of migration $\phi$ changes over time during some typical turns. For each turn, $\phi$ appears to vary linearly with time, thus featuring a near-constant rate of turning. Distributions of the turning rates are identical for turns of both signs, see Fig.~\ref{fig:rotor}(b) and inset.

The relationship between the total duration of a turn $t_{\rm turn}$ and the total angular change during it $\Psi=\phi(t_{\rm turn})-\phi(0)$ are displayed in Fig.~\ref{fig:rotor}(c). The ensemble of dots represents $N$=2458 turning events collected from the tracks, and the squares represent the mean of binned data, $\langle \Psi \rangle$. A clear linear dependence of $\langle \Psi \rangle$ on $t_{\rm turn}$ is found: $\langle \Psi \rangle= \Omega_{\rm turn} t_{\rm turn}$. Least-square linear fitting gives $\Omega_{\rm turn}=0.160\pm0.003$ rad/min, where the uncertainty represents the standard error. Meanwhile, we observe that the distribution of $\Psi$ becomes more dispersed for longer turns (larger $t_{\rm turn}$). This can be seen from the widths of the probability distributions of $\Psi$ in Fig.~\ref{fig:rotor}(c) inset: short ($t_{\rm turn}\leq8$ min, blue), intermediate ($t_{\rm turn}\in(8,16]$ min, purple), and long turning events ($t_{\rm turn}\in(16,32]$ min, red) have increasingly broadened distributions. In other words, longer turns have more variability around the expected angular change. Angular changes for runs are displayed in Fig.~\ref{fig:rotor}(d). Here $\Psi=\phi(t_{\rm run})-\phi(0)$ with $t_{\rm run}$ the duration of a single run. $\langle\Psi\rangle$ increases with $t_{\rm run}$ orders of magnitude more slowly than it does during turns. Quantitatively, the rate reads 0.006 rad/min and is $\sim$30 times smaller than $\Omega_{\rm turn}$, and can originate from diffusion (\App{B}). Similarly, the inset of Fig.~\ref{fig:rotor}(d) displays the probability distributions of $\Psi$ for short ($t_{\rm run}\leq30$ min, blue), intermediate [$t_{\rm run}\in(30,60]$ min, purple], and long runs [$t_{\rm run}\in(60,120]$ min, red]. The dispersity also increases over time but it increases only slightly, much slower than its counterparts during turns. 

It is intriguing to see that the cells, on an ensemble level, behave as noisy constant-rate rotors. We further resolve the noise level by computing the cells' angular auto-correlation functions (ACFs). Figure~\ref{fig:rotor}(e) displays the ensemble average ($\langle\,\cdot\,\rangle$) of single-cell ACF$\equiv\bm{n}(t') \cdot\bm{n}(0)$, with $\bm{n}(t')$ the unit vector of the cell's moving direction at time $t'$. The term $\bm{n}(t') \cdot\bm{n}(0)$ readily reduces to $\cos{(\Delta \phi(t'))}$ with $\Delta \phi(t') \equiv \phi(t')-\phi(0)$. We further assume that the noise is Gaussian and corresponds to a rotational diffusion coefficient of $D_r$. The Langevin equation for a noisy rotor with a persistent turning rate $\Omega$ reads $\dot{\phi}=\Omega + \zeta(t)$, with $\zeta(t)$ the noise that satisfies $\langle \zeta(0)\zeta(\tau)\rangle=2D_r\delta(\tau)$ and $\delta$ the Dirac delta function. For such rotors, their average ACF evolves as: 
\begin{equation}\label{eq:ACF}
    \langle\text{ACF}\rangle=\langle\cos{(\Delta \phi(t'))}\rangle =e^{-D_r t'} \cos{\Omega t'} 
\end{equation}
For small $\Omega$, that is, during runs, the ACF further reduces to $\langle\text{ACF}\rangle \approx e^{-D_r t'}$. The experimentally obtained ACFs for turns and runs follow these descriptions precisely, see Fig.~\ref{fig:rotor}(e). Fitting the ACFs of turns with \eqref{eq:ACF} gives $\Omega_{\rm turn}=0.160$ rad/min, agreeing with the value obtained by linearly fitting of $\langle\Psi\rangle$ as a function of $t_{\rm turn}$ [0.160 rad/min, Fig.~\ref{fig:rotor}(c)].
The rotational coefficients for turns and runs read respectively $D_{r,\rm{turn}}=0.031~ \rm{rad^2/min}$ and $D_{r,\rm{run}}=0.004~ \rm{rad^2/min}$. Notably, the angular noise is an order of magnitude lower for runs and shows why the distribution of $\Psi$ maintains nearly the same dispersity over time, as displayed in Fig.~\ref{fig:rotor}(d) inset.

The message that the cells are alternating between being persistent runners and constant-rate rotors is echoed by other statistics, i.e., the mean square angular displacement ($\langle\Delta\phi^2\rangle$, MSAD) and mean square displacement ($\langle\Delta r^2\rangle$, MSD), Fig.~\ref{fig:rotor}(f). Angular changes (MSAD) accumulate near-ballistically over time for turns. For runs, notably, on time scales shorter than the events' characteristic duration
$\tau_{\rm run}$ ($\sim$30 min), MSAD only increases sub-diffusively, indicating a strong suppression of angular noise during runs.
Only on time scales longer than $\tau_{\rm run}$, the angular changes start to build up diffusively. For spatial displacement, the trend is the opposite. MSD is ballistic for runs whereas (sub-)diffusive for turns.

We also examine whether single cells have their own characteristic turning rate or a preference for turning sign (\App{C}). We observe that the turning rates of a single cell during its consecutive turns are highly randomized, and these rates are as scattered as those observed across an ensemble of cells [Fig.~\ref{fig:chirality}(a)]. The data does not support that single cells have characteristic turning rates.   Also, they neither display a preferred sign of turning [Fig.~\ref{fig:chirality}(b)].

\subsection{Mechanotaxis achieved by biasing the sign of turns}

If a cell migrates only with two modes, namely, runs along an unchanged direction and turns at a constant rate, how does it perform targeted motion? An intuitive answer is that the cell can bias its direction of turning towards the stimuli to approach them. We now test this hypothesis.

\begin{figure}[htbp]
    \includegraphics[width=0.48\textwidth]{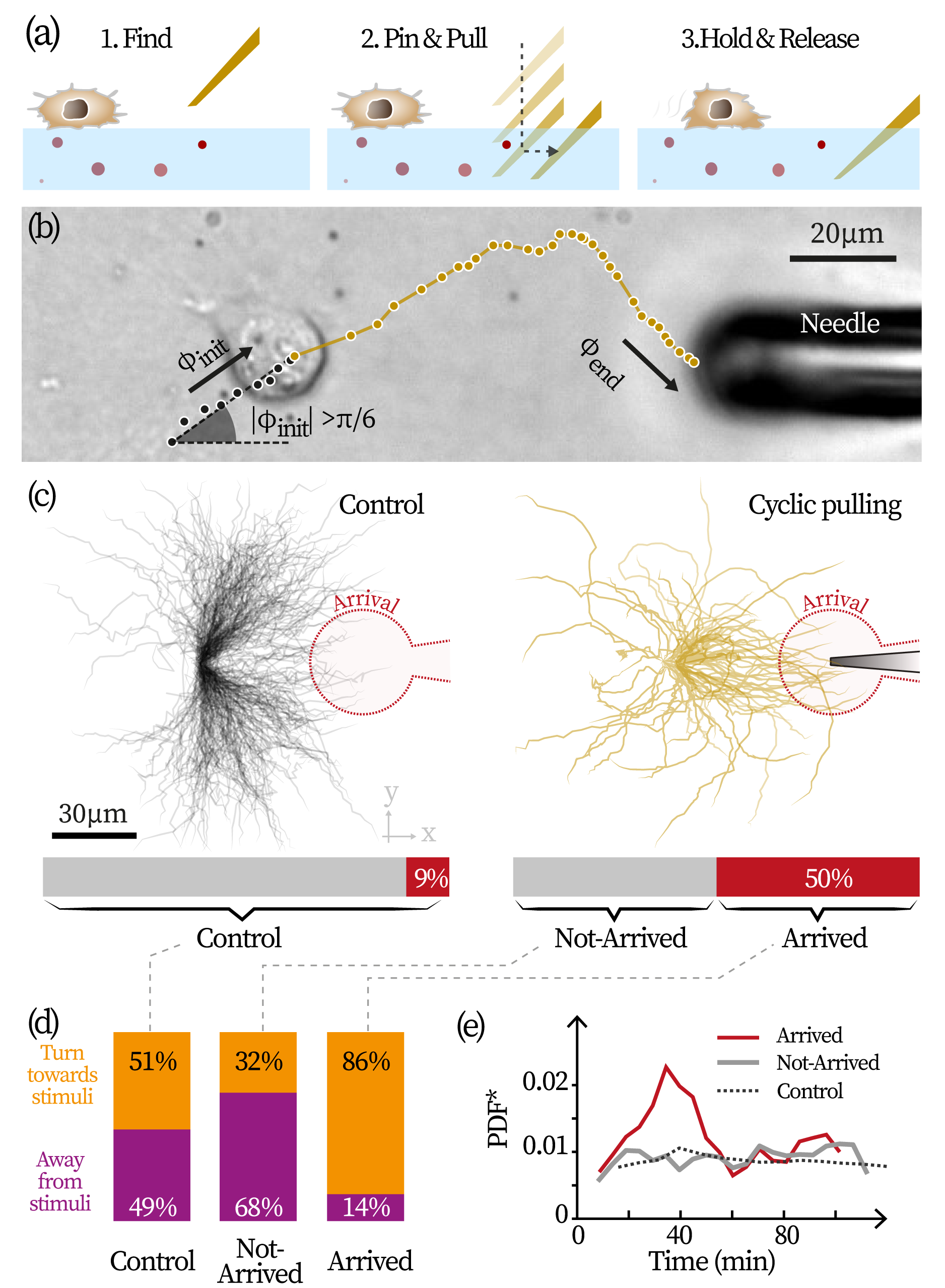}
    \caption{Mechanotaxis of cells. (a) Experimental scheme for exerting cyclic pulling. (b) A typical eligible track of an attracted cell. The black segment represents pre-stimuli observation. (c) Visualization of eligible tracks in the control group (black, $N$=293) and tracks in mechanotaxis assays (yellow, $N$=93). The fraction of cells arrived at the red-shaded areas are reported by bottom bars. (d) Directional distributions and (e) temporal distributions of turns in different cell groups. The distribution of arrived cells in (e) is adjusted to account for the tracks' varying duration, see Methods for detail.}
    \label{fig:taxis}
\vspace{-3mm}
\end{figure}

Following a similar protocol as Ref.~\cite{Zhang2023microglia}, cyclic pulling is applied to the substrate on which the cells are deposited, serving as mechanical cues. 
In short, micropipettes were inserted into the collagen substrate 60~\um to the right of the cell. They are then used to pull the substrates further rightwards, sustaining the stretch for some time. Lastly, the substrate is released and the pipette goes back to the initial location to start another cycle, see   Fig.~\ref{fig:taxis}(a) for an illustration and see Methods for details.

We examine single-cell trajectories in response to the mechanical stimuli. Before insertion of the micropipette, we ensure that the candidate cell is motile and its moving direction does not point directly to the stimuli [$\phi(0) \notin [-\frac{\pi}{6}, \frac{\pi}{6}]$]. Moreover, because we observe that cells originally moving leftwards [$\phi(0) \in [\pm\frac{\pi}{2}, \pm\pi$]] do not get attracted to move to the pipette, thus we also exclude this angular range. A typical eligible track is displayed in Fig.~\ref{fig:taxis}(b), with the black fraction representing the cell's pre-stimuli migration and the yellow segment its motion after the stimuli is applied. Note that $\phi(0)$ is computed based on the cell's pre-stimuli trajectory. See also SV.~2 for an attracted cell. In total, $N=93$ motile cells whose initial angles are in the desired range are gathered for this study. To comprise a control group, $N$=293 eligible tracks are collected. These cells are motile, have proper initial directions, and migrate with no pipette inserted into the substrate. The two groups of tracks are displayed in Fig.~\ref{fig:taxis}(c).

A cell is considered to be attracted if it has arrived at the pipette within 120 min, or its distance to the needle tip at the end of 120 min is smaller than 20 \um. The region of arrival is visualized by the red-shaded area in Fig.~\ref{fig:taxis}(c). Compared to the control group where only 9\% of cells ($N$=26 out of 293) arrived at the stimuli, $\sim50\%$ cells ($N$=47 out of 93) under cyclic pulling of the substrate are attracted [the bars at the bottom of Fig.~\ref{fig:taxis}(c)]. Since this region has no actual difference from elsewhere for the control group, the arrival fraction should be interpreted as the baseline probability for an eligible cell to end up in such an area of complex geometry.

To resolve the direction of turning of cells under stimuli, we pool all detected turning events and divide them into three groups: the turns taken by the arrived cells, by the not-arrived cells, and additionally, those by the cells in control group (as a whole). A turn is considered helping the cell get closer to the stimuli if it decreases $\theta\equiv \arccos(\bm{n}\cdot\bm{n}_{c\rightarrow p})$, which denotes the included angle between the unit vector along a cell's moving direction, $\bm{n}$, and the unit vector pointing from the cell's location at the moment of turning, to the pipette tip, $\bm{n}_{c \rightarrow p}$.

Figure~\ref{fig:taxis}(d) displays the fractions of the turns that help cells move towards the stimuli (orange) or away from the stimuli (purple). While turns in the control group does not favor either sign, 
among the arrived cells, the turning sign is predominantly biased to help the cell migrate towards the stimuli (86\%). To confirm whether the turning sign is biased due to mechanotactic responses, we benchmark the temporal distribution of the turns in the arrived group against the control group, see Fig.~\ref{fig:taxis}(e). In the control group, turns take place at a constant rate, featuring a flat line in the turns' probability distribution over time. In sharp contrast, among the arrived cells, the probability peaks during the first $\sim$50 min after the onset of stimuli (at t=0 min), meaning that the cells have become more likely to turn. Such responsiveness indicates that the majority of the arrived cells are actively engaged in mechanotaxis. In other words, the probability rates of turning show that the temporal symmetry is maintained in the control group, whereas it is broken in the arrived group by the introduction of external stimuli.

It is noteworthy that the cells under stimuli do not always respond to stimuli. The temporal distribution of turn of not-arrived cells shows no response to stimuli, remaining essentially the same as the control group, see Figs.~\ref{fig:taxis}(d) and \ref{fig:taxis}(e). Meanwhile, their signs of turns are slightly biased away from the stimuli, possibly because some non-responsive cells that happen to arrived at the stimuli have been excluded.

To summarize, our data show that, upon exposure to mechanical stimuli, a considerable fraction of MCF-10A cells are attracted, although not all are responsive. For those responsive cells, their mechanotaxis is underpinned by biasing the overall direction of turning towards the stimuli.

\subsection{Migration of mesenchymal MCF-10A and NIH-3T3 cells}

\begin{figure}[htbp]
    \includegraphics[width=0.48\textwidth]{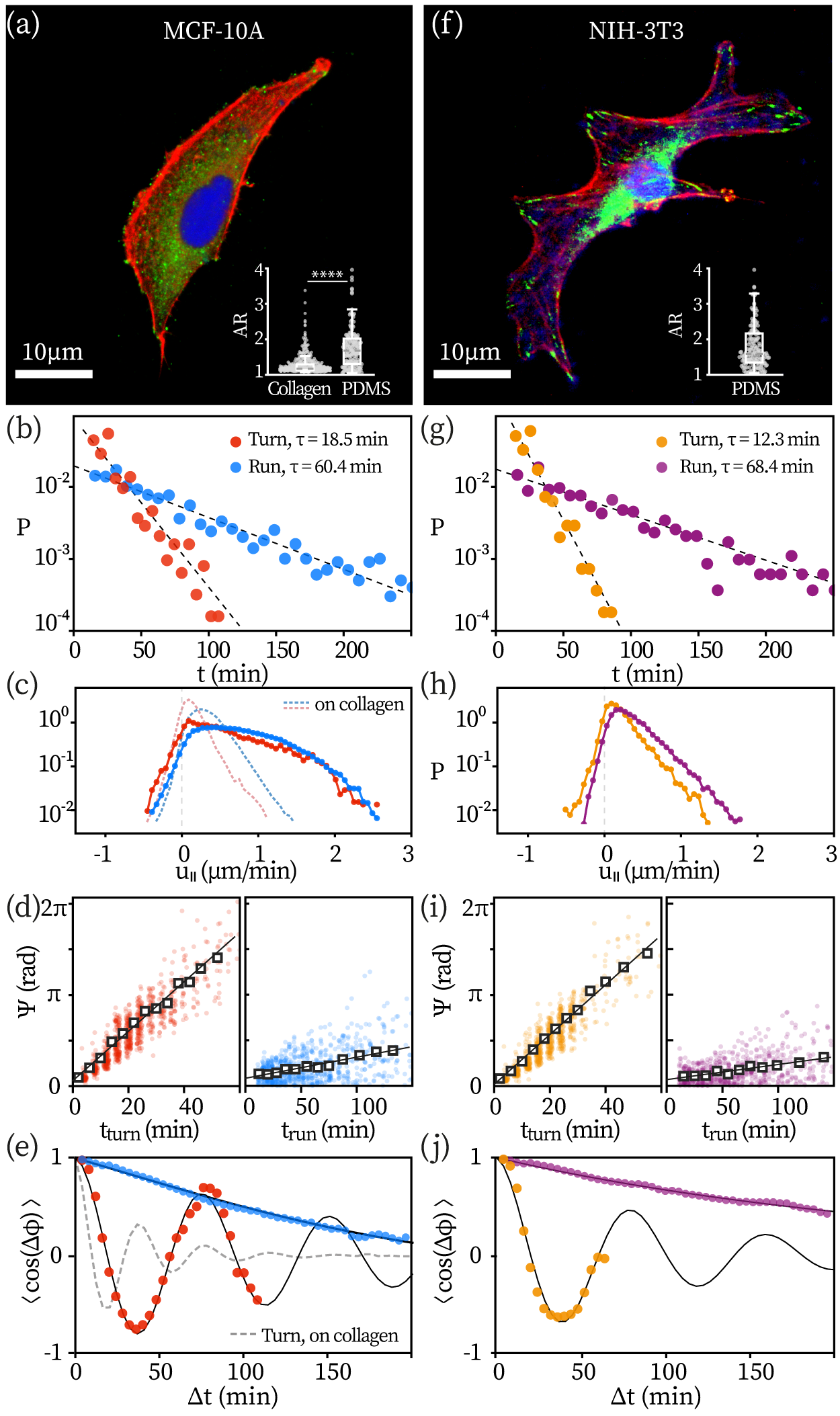}
    \caption{Runner-rotor motility in other cells. (a) Fluorescent microscopy of MCF-10A cell examined on PDMS substrates. Inset: comparison of aspect ratio (AR) between cells examined on different substrates. ****:p<0.001, Kruskal-Wallis test, one-way ANOVA. Distributions of duration (b) and velocity along the track direction (c) for turns and runs. (d) Total change in direction, $\Psi$, as a function of a state's duration. (e) Angular auto-correlation function for both states. Data are fitted exponentially in b (dashed line), linearly in d (solid line), and by \eqref{eq:ACF} in e (solid line). Panel (f)-(j) respectively show information of NIH-3T3 cells in the same form as (a)-(e).} \label{fig:mesenchymal}
  \vspace{-3mm}
\end{figure}

Results shown so far indicate MCF-10A cells on collagen ($\sim10$ Pa), which are predominantly in the amoeboid morphology, switch between being persistent runners and constant-rate rotors. Cells in such morphology 
tend to migrate without mature focal adhesions and stress fibers~\cite{lammermann2009mechanical}. On different substrates, the cells' motility and morphology can be different~\cite{Friedl2010plasticity}. We therefore examine MCF-10A on stiffer substrates (PDMS, 750~kpa).
In this case, the cells migrate in a mesenchymal fashion:
they adopt an elongated, spindle-like shape and exert traction on their substrates via focal adhesions associated with actin-rich protrusions, such as lamellipodia or filopodia~\cite{Pollard2003cellular}. We find that the runner-rotor model still precisely captures the cells' migration kinematics.

In total, $N$=318 motile MCF-10A cells are examined on PDMS. The fluorescent image of a typical MCF-10A in mesenchymal mode is displayed in Fig.~\ref{fig:mesenchymal}(a). The cells' different morphologies can be seen straightforwardly from the significant difference in the cells' aspect ratio (AR) [inset of Fig.~\ref{fig:mesenchymal}(a), p<0.001 by Kruskal-Wallis test, one-way ANOVA]. We define AR as the ratio of the longest side to the shortest side of the smallest bounding box of the cell's outline. Cells examined on collagen have concentrated around $\rm{AR}=1$, while on PDMS, they have AR distributed between 1 and 2 in a nearly uniform fashion.

The mesenchymal cells also have exponentially distributed state intervals for runs and turns, Fig.~\ref{fig:mesenchymal}(b). In these cells, a run persists approximately twice as long as in their amoeboid counterparts. 
Mesenchymal MCF-10A cells also run faster: $\overline{V}_{\rm run}=0.76\pm0.32$ \umm (mean$\pm$std.) versus $0.39\pm0.16$ \umm for cells on collagen. Figure~\ref{fig:mesenchymal}(c) further resolves the distribution of the velocity component parallel to the cell's moving direction, during both turns (red solid line) and runs (blue solid line). Both traces reach further to the positive values compared to the amoeboid data (dashed lines), indicating much enhanced motility.

Nevertheless, on the ensemble level, cells still turn at a near-constant rate ($\Omega_{\rm turn}=0.090\pm0.003$ rad/min) and is an order of magnitude faster than they do in runs (0.007 rad/min), see the left and right panels in Fig.~\ref{fig:mesenchymal}(d) respectively. Fitting the cells' ACF with \eqref{eq:ACF} [solid lines in Fig.~\ref{fig:mesenchymal}(e)] gives essentially the same turning rate. For the sake of space and clarity, comprehensive fitting results are summarized in Table \ref{table:fitting}. Notably, the mesenchymal cells turn with much stronger coherence than the amoeboid cells, which can be directly seen from Fig.~\ref{fig:mesenchymal}(e), that the oscillatory pattern of ACF lasts much longer than for the cells examined on collagen (dashed line). Quantitatively, the turning noise in mesenchymal cells ($D_{r,\rm{turn}}=0.005\ \rm{min}^{-1}$) amounts to less than 20\% of the noise in amoeboid cells ($0.031\ \rm{min}^{-1}$).

Clearly, the runner-rotor model applies precisely for MCF-10A cells in mesenchymal state. These cells with stronger morphological polarity [Fig.~\ref{fig:mesenchymal}(a) inset] are found to run more persistently and faster [Figs.~\ref{fig:mesenchymal}(b) and \ref{fig:mesenchymal}(c)]. Meanwhile, they are subjected to much less rotational noise during turns [Fig.~\ref{fig:mesenchymal}(e)]. This is in line with the UCSP model~\cite{Maiuri2015}, that cell polarity, speed, and persistence are positively correlated.

Does the runner-rotor kinematics observed in epithelial cells (MCF-10A) applies for other cells? We examined the well-characterized fibroblast cell (NIH-3T3, $N$=280) cells on the same PDMS substrate. The cells display a morphology similar to the MCF-10A cells, Fig.~\ref{fig:mesenchymal}(f) and inset. The state intervals of NIH-3T3 again follow two exponential distributions, in line with previous study~\cite{Begemann2019}, Fig.~\ref{fig:mesenchymal}(g). Results on the cells' motility, turning rates, and the coherence of turning are presented in Figs.~\ref{fig:mesenchymal}(h), \ref{fig:mesenchymal}(i), and \ref{fig:mesenchymal}(j) respectively. These results are analogous to those obtained from MCF-10A cells, see Table \ref{table:fitting} for the parameters of fitting. The precision of applying the current framework to another type of cell evidences the general applicability of the runner-rotor kinematics. Typical migration of MCF-10A and NIH-3T3 cells can be found in SV.~3 and 4 respectively.

\begin{table*}
\centering
\begin{tabular}{ccc|cccccc}
Cell & Number &Substrate & $\tau_{\rm turn}$& $\tau_{\rm run}$ & $\Omega_{\rm turn}$& $D_{r,\rm turn}$&$D_{r,\rm run}$\\
 & $N$= & ~ &(min)&(min)&(rad/min)&($\rm{min}^{-1}$)&($\rm{min}^{-1}$)\\
\midrule
MCF-10A & 396 &collagen & 8.2 & 29.9 & 0.160\,(0.160) &0.031&0.005\\
MCF-10A & 318 & PDMS &18.5 & 60.4 & 0.083\,(0.090) & 0.006&0.004\\
NIH-3T3 & 280 &PDMS &12.3 & 68.4 & 0.082\,(0.086) & 0.011&0.005\\
\end{tabular}
\vspace{3mm}
\caption{Kinematic properties for different cells. $\tau_{\rm run,turn}$ is obtained from exponential fitting the state interval distributions; $\Omega_{\rm turn}$ and $D_{r, \rm run,turn}$ are obtained from fitting the ACF with \eqref{eq:ACF}. The values in parenthesis $(\cdot)$ represent results of linear fitting of $\langle \Psi \rangle$ with $t_{\rm run, turn}$.}
\label{table:fitting}
\vspace{0mm}
\end{table*}

\subsection{UCSP revisited and the coupling between turning rate and duration}
Conventionally, a cell's persistence is measured often in an \textit{ad hoc} fashion. In some cases, persistence is experimentally measured as the time needed for a cell to turn 90\degree~\cite{Maiuri2015, Jerison2020}; or the total time span where frame-to-frame turning angle is less than 30\degree~\cite{Werner2019,Begemann2019}; or obtained by fitting tracks ~\cite{Alessandro2017,Shaebani2020} or MSDs to models~\cite{Leineweber2023}. While these practices have generated valuable insights in their respective systems, the variability makes cross-cell comparison difficult. In this section, we show the insights brought about by the runner-rotor kinematics and its implications for the UCSP model. 

Given a cell is constantly switching between running and turning, the practice of reporting persistence as the time corresponds to 90\degree-change in migration direction would result in large uncertainty, because the reported value strongly depends on where the measurement starts. Moreover, by modeling runner-rotor kinematics, we find that, when persistence is reported as such, the typical exponential dependence of persistence on the speed~\cite{Maiuri2015}, which are used to justify UCSP, does not suffice this purpose. In fact, the exponential dependence may manifest even without any coupling between persistence and speed, Fig.~\ref{fig:challenge}. Detailed information can be found in \App{D}.

\begin{figure}[htbp]
    \includegraphics[width=0.48\textwidth]{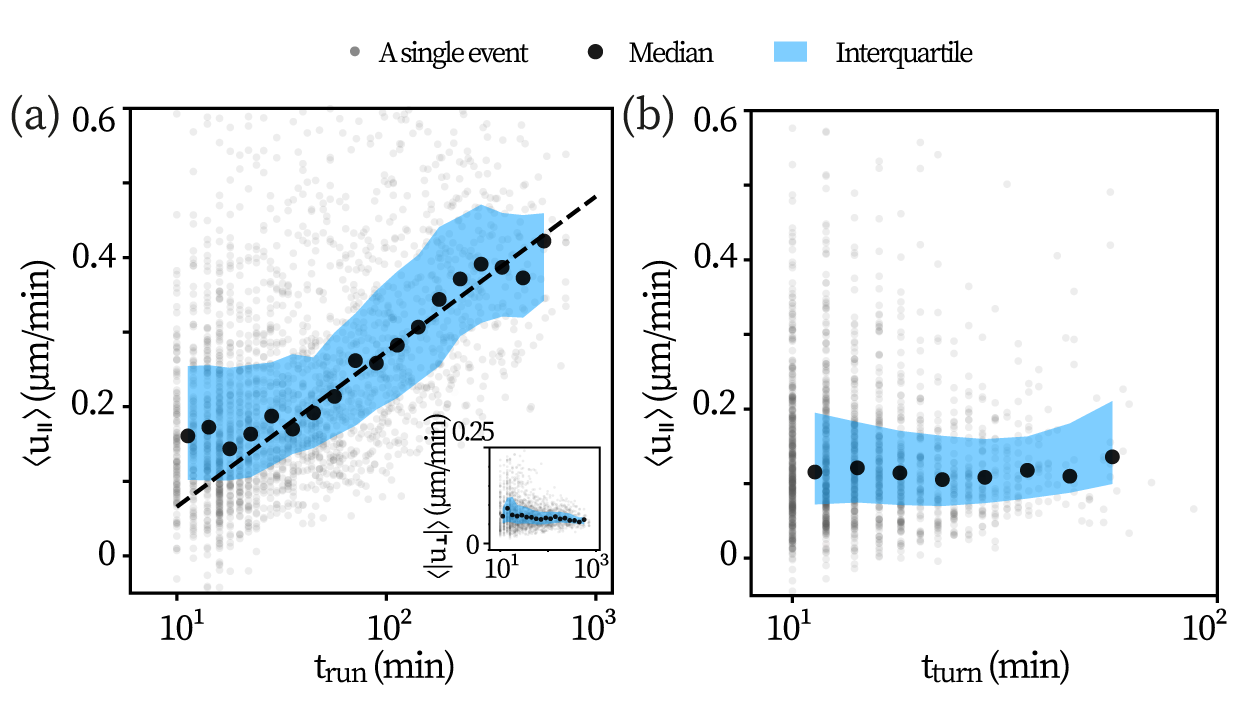}
    \caption{UCSP examined with cell migration dichotomized as runs and turns. (a) Correlation between cell speed and the duration of runs ($N$=2413) and (b) turns ($N$=1657) that are longer than 10 min.} \label{fig:positiveFeedback}
  \vspace{-3mm}
\end{figure}

On the other hand, the runner-rotor model gives an unambiguous definition of persistence time: the duration of a run $t_{\rm run}$. With this refined definition, we re-examine if the cells' speed and persistence are coupled. We again decompose the cells' instantaneous velocity as $u_{\perp}$ and $u_{\para}$. Figure~\ref{fig:positiveFeedback}(a) displays the results of runs collected from MCF-10A cells on collagen. Each dot in the background represents a run event, while the black circles and the shading represent respectively the median and interquartile of the data binned by $t_{\rm run}$.
The mean effective component for forward-motion, $\langle u_{\para} \rangle$, is found to be strongly coupled with the total duration of run $t_{\rm run}$, see Fig.~\ref{fig:positiveFeedback}(a); whereas the perpendicular component, $\langle \left| u_{\perp} \right| \rangle$ is not coupled with $t_{\rm run}$, see Fig.~\ref{fig:positiveFeedback}(a) inset. Data for MCF-10A and NIH-3T3 cells tested on PDMS substrates show qualitatively the same results. Altogether, these results provide a solid and nuanced support to the UCSP model.

\begin{figure}[htbp]
    \includegraphics[width=0.48\textwidth]{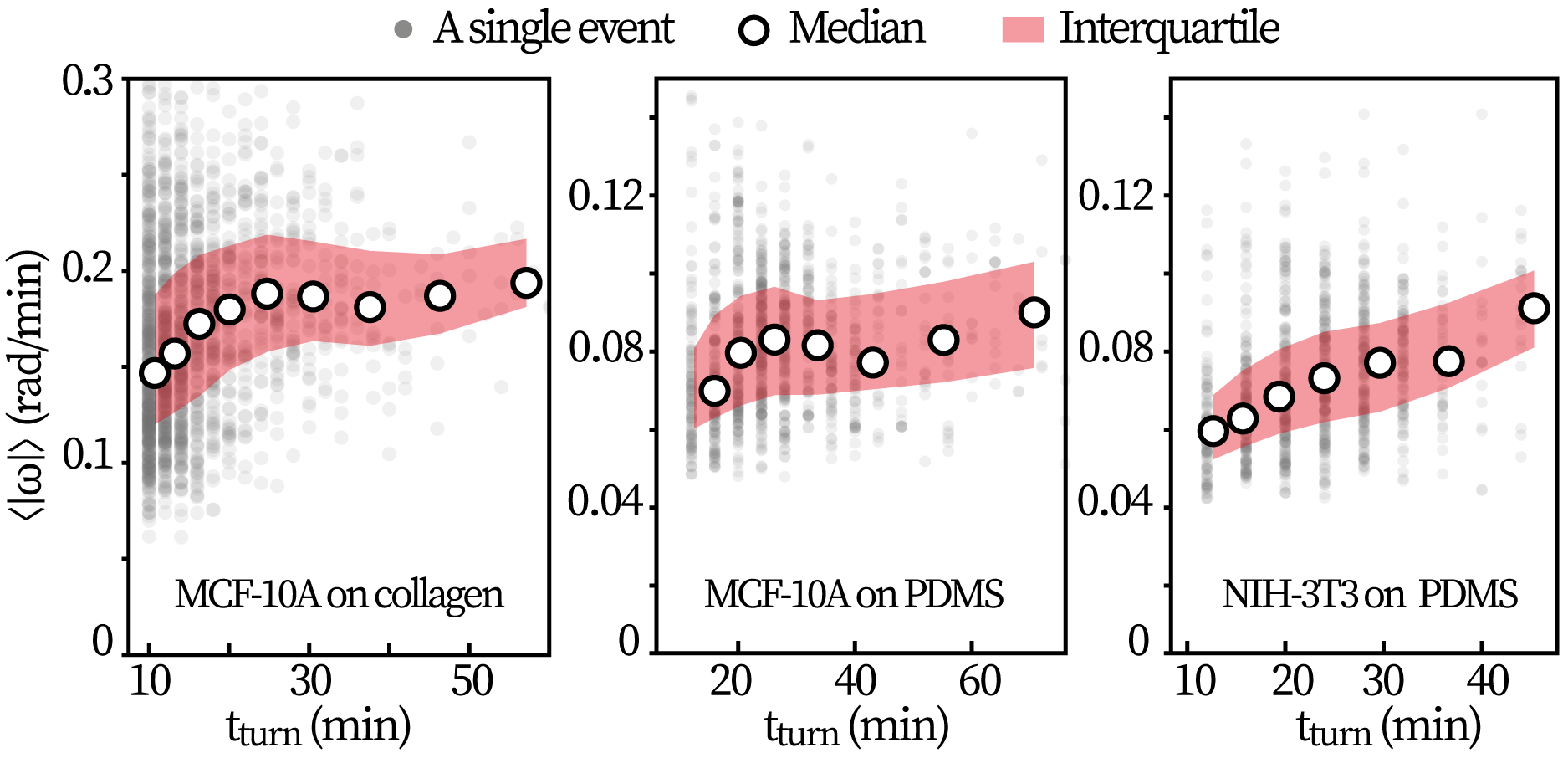}
    \caption{Positive correlation between the rate and duration of turns. Panels correspond respectively to different datasets as annotated. Note that the $x$-axes are linear.} \label{fig:positiveFeedback_omega}
  \vspace{-3mm}
\end{figure}

Naturally, is is also of interest to see if migration speed is also correlated with the duration of turns. 
Figure~\ref{fig:positiveFeedback}(b) shows the relationship between $\langle u_{\para} \rangle$ and the total duration of turn ($t_{\rm turn}$) from MCF-10A cells on collagen. How fast a cell moves during turns appears to have no correlation with how long this turn would last. 
Next, instead of analyzing how $t_{\rm turn}$ is correlated to the perpendicular speed component $u_{\perp}$, which fluctuates strongly over time due to the cells' morphological changes, it is more instructive to directly analyze the correlation between the angular velocity ($\omega$) and the duration $t_{\rm turn}$ of a turn.

In a previous study using fish keratocyte~\cite{Allen2020}, it is suggested that the turning rate and its duration are mutually enhancing factors. Indeed, although turning rates within a dataset only vary within a small range (20\%-30\%), yet a clear positive dependence between the mean turning rate, $\langle \left|\omega\right|\rangle$ and $t_{\rm turn}$ is observed, see Fig.~\ref{fig:positiveFeedback_omega}. In short, the cells that turn faster also tend to turn for a longer period of time. 

\section{Discussion}
In this work, we demonstrate a framework to assess how adherent cells turn. On an ensemble level, the cells' turning kinematics (under a given condition) feature a nearly constant turning rate but stochastic turning signs. When cells undertake targeted migration, they bias the signs towards the stimuli. Separating turns and runs with our protocol and analyzing the states respectively brings new insights. The analysis of runs provides a solid and nuanced support for the UCSP model~\cite{Maiuri2015}; while the analysis of turns evidences the hypothesized positive feedback between turning rate and duration~\cite{Allen2020}. Lastly, while adherent eukaryotic cells clearly exhibit runner-rotor kinematics, non-adherent cells such as immune cells migrate with run-and-tumble kinematics~\cite{Zhang2023microglia} --- their turns are purely diffusion-driven. This distinction highlights the role of a eukaryotic cell's mechanical interactions with the substrate in determining its turning mechanism.

Two remarks must be made about the claim of 'turning at a constant rate'. (1) One should not confuse turning at a constant rate with circling, where 'circling' describes a track's turning fraction resembling a circular arc. This is because circling requires not only constant-rate turning but also a near-constant speed of forward motion. However, the propagation speed $u_{\para}$ of our cells varies strongly over time, which constitutes the probability distributions displayed in Figs.~\ref{fig:twoState}(c), \ref{fig:mesenchymal}(c), and \ref{fig:mesenchymal}(h). With Fig.~\ref{fig:demo_track} in \App{A}, we further demonstrate this point with typical tracks with long turning events: the appearance of these turns is disparate from a circle or an arc. Nevertheless, circling behaviors do exist. They have been reported in fish keratocytes, which migrate $\sim10\times$ faster than the cells studied here and have distinct morphologies~\cite{Allen2020}. (2) On the scale of single turns, single cells, and the ensemble, 'turning at a constant rate' means differently. Per single turn, we observe that there exists a constant rate that does not vary over time, Fig.~\ref{fig:rotor}(a). Per single cell, the rates for each of its turns are stochastically determined, from approximately the same probability distribution as the ensemble, Fig.~\ref{fig:chirality}(a). On the ensemble level, the constant-rate turning manifests as a linear trend between the most probable angular change ($\langle \Psi \rangle$) of a turn and its duration ($t_{\rm turn}$), see Figs.~\ref{fig:rotor}(c) \ref{fig:mesenchymal}(d), and \ref{fig:mesenchymal}(i).

Considering how much the cell's properties (e.g. morphology, speed, size) vary from one to another within a population, the existence of a most probable rate of turning is surprising and has crucial implications for modeling the kinematics of adherent cell migration. Previously, the turning dynamics are typically understood as a diffusive process~\cite{Maiuri2015,Alessandro2017}, where $\dot \phi(t) \sim \zeta(t)$, with $\zeta$ a zero-mean Gaussian noises. However, our results indicate that turnings should be described as:
\begin{equation}\label{eq:turningDynamics}
    \dot \phi = \Omega + \zeta(t),
\end{equation}
where $\Omega$ accounts for the deterministic turning. The two descriptions of turning are fundamentally different.

Lastly, to explain the kinematics described in \eqref{eq:turningDynamics}, we combine the insights from Ref~.\cite{Jiang2023} and Ref.\cite{Allen2020}. In runs, only actin filaments perpendicular to the cell membrane would polymerize fast enough to stay in contact with the forward-moving membrane. The actin flow and the membrane motion reinforce each other, such that the direction of run dominates the actin flows. However, at a certain point, when the actin flow is not strong enough to push the cell forward, the direction of actin filament polymerization would bifurcate. At this instant, the likelihood of both leftwards and rightwards actin flow - and consequently, the cell's turning left or right - is symmetric. However, due to positive feedback between polymerization and membrane deformation and the limited source of actins, the left-right symmetry will break: one turning direction will out-compete the other and lead the cells to turn~\cite{Jiang2023}. After the symmetry-break, accumulation of myosin-II at the outer and rear side and the asymmetric actin flow start mutually enhancing each other~\cite{Allen2020}. This positive feedback probably sustains the turning at a near constant rate. However, what sets the mean turning rate over a population of cells remains to be clarified. Future experiments are called for to elucidate this question.\\

\section*{Acknowledgments} 
We thank Hui Li and Mingcheng Yang for helpful discussions. This work is supported by the National Key Research and Development Program of China (2022YFA1405002); the National Natural Science Foundation of China (NSFC) (Grant Nos. 12204525, 12325405, 12090054); the Youth Innovation Promotion Association of CAS (No. 2021007). 

\section*{Author contributions} 
Y.Z. and X.Y. performed experiments; Q.F. and F.Y. designed experiments; B.Z. and D.W. performed simulations; D.W. analyzed data and wrote the manuscript. D.W. and F.Y. conceived and supervised the project. All authors reviewed the manuscript.  

\renewcommand{\thefigure}{S\arabic{figure}}
\setcounter{figure}{0}
\setcounter{section}{0}

\appendix
\section{Detect turns by high turning rates}

Initially, we attempted to distinguish the states of run and turn by the cells' instantaneous speed, with similar algorithms used in \cite{Zhang2023microglia}. The attempt was not successful. In hindsight, two major reasons are: (1) The cells' instantaneous speed during the two states is not clearly discernible, see for example, the speed distributions in Figs.~\ref{fig:mesenchymal}(c) and \ref{fig:mesenchymal}(h); (2) While running, the retraction of lamellipodia results in intermittent slow-speed periods. The speed-based algorithms tend to confuse these intermittent slow phases of running with turning.   

However, we find that runs and turns can be effectively separated by thresholding the cells' turning rate. To begin with, the cell's direction of motion $\phi(t)$ is computed from its smoothed track (moving-window smoothing of $t_{
\rm win}$), $\phi(t)=\arctan{\left(\frac{y(t+t_{\rm win}/2)-y(t-t_{\rm win}/2)}{x(t+t_{\rm win}/2)-x(t-t_{\rm win}/2)}\right)}$. To extract a meaningful net migration direction of the cell, $t_{\rm win}$ is chosen to meet three criteria: (1) it should be longer than the cell's fast contraction timescale to avoid artifacts from rapid boundary deformations ($\sim$10 min, see also~\cite{Begemann2019}); (2) the cell's net displacement over $t_{\rm win}$ (estimated from the cell's speed) should significantly exceed image detection errors ($\sim$3 pixels or $\sim$2 $\mu$m); and (3) $t_{\rm win}$ should be as short as possible to maximize the temporal resolution for identifying the run/turn states.

Consecutive frames with turning rate higher than $\Omega_c$ are marked as turns. Same type of events separated by a 1-frame gap are merged (gap-closing). Thereafter, the consecutive frames wherein the cell turns at a rate higher than the threshold $\Omega_c$ is considered as a single turning event. Finally, one needs to properly justify a threshold $\Omega_c$.

Empirically, one may apply $\Omega_c$ to tens of tracks simultaneously and decide by visual observation if turning events are properly marked. Nevertheless, when the number of tracks increase and different cell lines are employed, a formalized protocol is called for. We develop a semi-automated protocol that gives $\Omega_c$ close to the empirical values determined by human users. In the following, the protocol is demonstrated with data obtained from MCF-10A cells examined on collagen ($t_{\rm win}$=7 frames, $\sim$12 min). For MCF-10A or NIH-3T3 cells examined on PDMS (4 min/frame), we keep using $t_{\rm win}$=7 frames$\sim$24 min. Using $t_{\rm win}$ of 5 or 9 frames does not affect the conclusions of this work but will (1) result in different values of $\Omega_c$ and (2) induce 10-20\% quantitatively changes in the reported values of $\tau_{\rm run, turn}$.

\begin{figure}[htbp]
\includegraphics[width=0.48\textwidth]{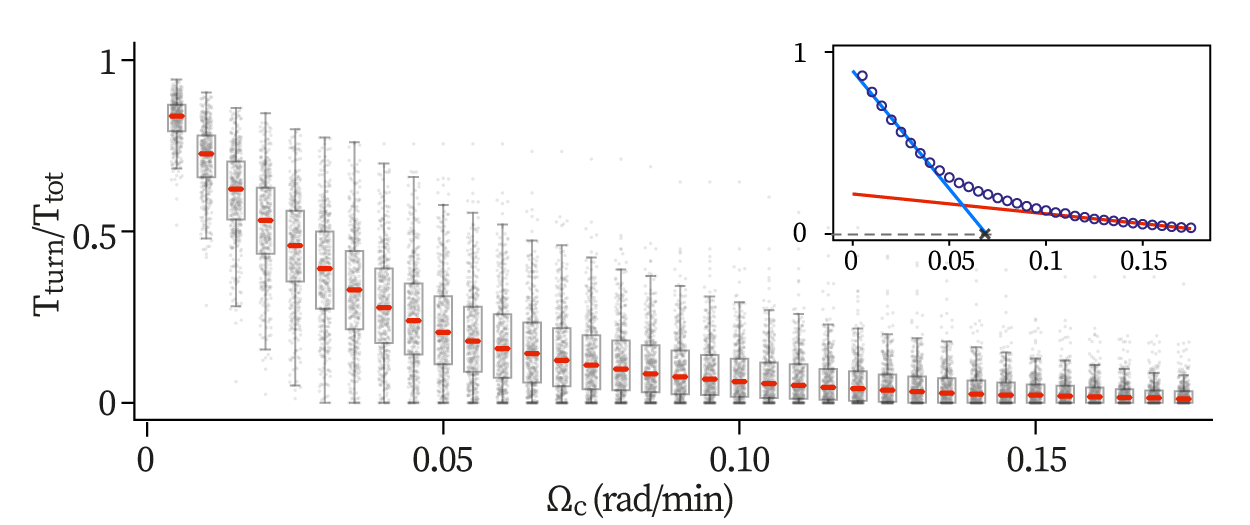}
    \caption{Determine $\Omega_c$ by the ensemble behavior of the time fraction of turning $T_{\rm turn}/T_{\rm tot}$. Each dot represent the time fraction of a single track. The boxplots mark the median over the tracks (thick horizontal line), the interquartile (black box), and 95-/5-percentile (top/bottom caps) of the data. Inset: median values with two linear extrapolations of the data in the left and right extremes.} \label{fig:omega_threshold}
  \vspace{-3mm}
\end{figure}

We scan $\Omega_c$ and observe how the time fraction of turns, $T_{\rm turn}/T_{\rm tot}$, evolves on the ensemble level as a function of $\Omega_c$, with $T_{\rm turn}$ the total time of turns within a single track and $T_{\rm tot}$ the duration of that track, see Fig.~\ref{fig:omega_threshold}. Under increasing $\Omega_c$, the median value of $T_{\rm turn}/T_{\rm tot}$ drops fast initially and then slows down, see the blue and red line in Fig.~\ref{fig:omega_threshold} inset. It initially drops fast as a large number of mislabelled turns due to fluctuations in $\phi(t)$ are filtered out. Further increasing $\Omega_c$, actual turns that have relatively low turning rates start to be disregarded, and the decrease of $T_{\rm turn}/T_{\rm tot}$ slows down. Therefore, the crossover between the fast and the slowly decreasing regime in $T_{\rm turn}/T_{\rm tot}$ can be used to determine the proper $\Omega_c$. We find that the $x$-intersection of the of linear extrapolation of the fluctuation-dominated regime (left side) agrees well with $\Omega_c$ obtained empirically~\footnote{ The curve of $T_{\rm turn}/T_{\rm tot}$ can also be fitted well by a mixture of two exponential distributions, whose characteristic scales help determine $\Omega_c$. However, fitting with the mixed exponential functions is more sensitive to small variation in data. It is thus deemed less robust and not employed.}.

\begin{figure}[htbp]
    \includegraphics[width=0.48\textwidth]{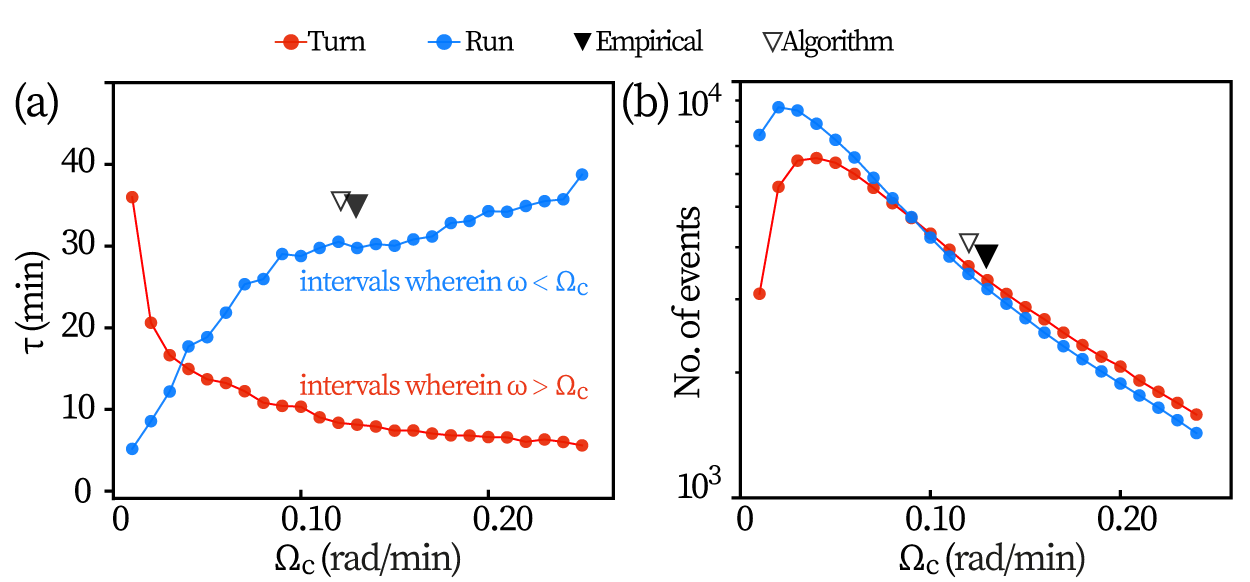}
    \caption{Different sensitivity of the two states to the choice of $\Omega_c$. (a) The mean duration of the so-regarded runs and turns, $\tau_{\rm run, turn}$ and (b) the total number of events as a function of $\Omega_c$. The thresholds computed by algorithm (Fig.~\ref{fig:omega_threshold}) and determined by empirical observation of human user are marked by triangles.}\label{fig:tauByOmegaC}
  \vspace{-3mm}
\end{figure}

\begin{figure}[htbp]
    \includegraphics[width=0.48\textwidth]{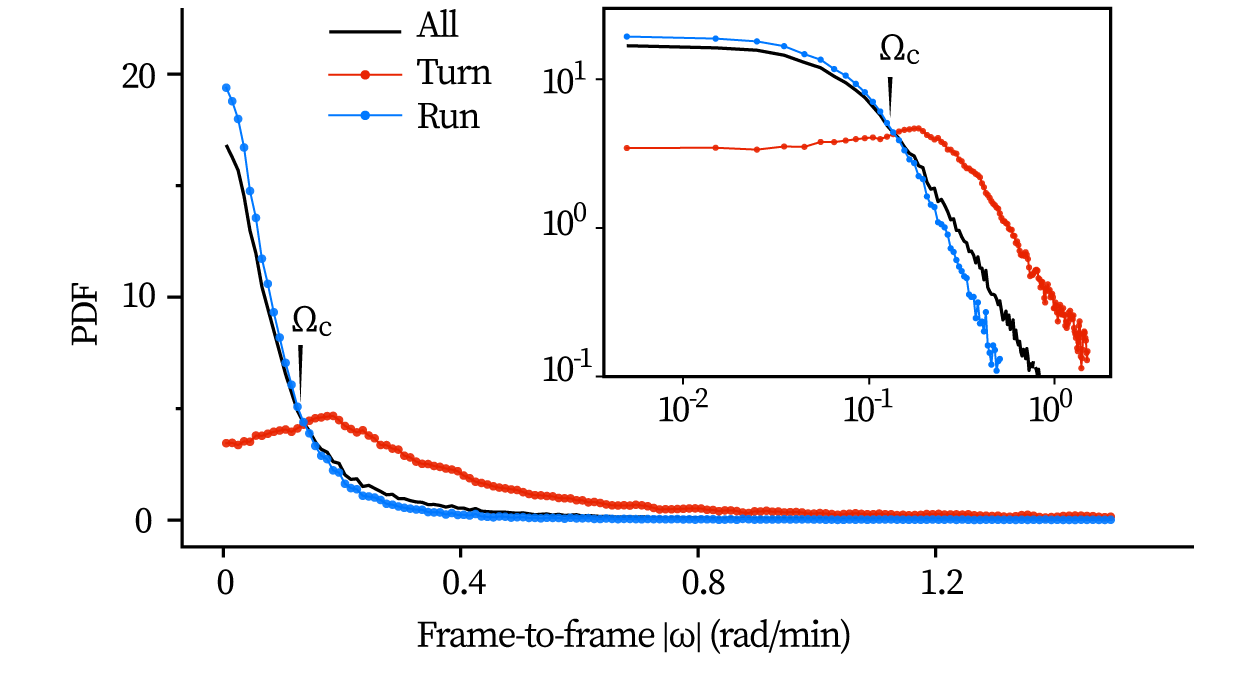}
    \caption{Distribution of frame-to-frame turning rate before and after distinguishing turns from runs. Inset: same data in log-log scale.} \label{fig:omega_distribution}
  \vspace{-3mm}
\end{figure}

Examination of the behaviors of the so-regarded 'runs' and 'turns' under different choices of $\Omega_c$ can also be used to justify the algorithm's choice. In Fig.~\ref{fig:tauByOmegaC}(a) we display the mean duration of the events $\tau$ in which the turning rate $\omega<\Omega_c$ (the so-regarded runs, blue) or $\omega>\Omega_c$ (the so-regarded turns, red). The mean duration of runs $\tau_{\rm run}$ displays clear saturation-like behavior. And we see that the threshold computed by algorithm (empty triangle) and that determined by empirically (solid triangle), being close to each other, locate both after the curve saturates. Figure~\ref{fig:tauByOmegaC}(b) display the number of so-regarded runs and turns under different $\Omega_c$. Events at the begin and the end of each track are discarded, as well as extremely short events. The threshold of choices (triangles) correspond to the entering of the regime where $N_{\rm turn} \gtrsim N_{\rm run}$. The number of turns is slightly higher because runs are more likely to be at the end and be excluded.

In Fig.~\ref{fig:omega_distribution}, we display the distribution of frame-to-frame turning rate. After runs and turns are separated, the distribution for turns peaks around $\Omega_{\rm turn}\approx 0.16$ rad/min. One should note that, because turns only take up $\sim$20\% of the total time, such a peak will not be discernible in the distribution before the states are isolated.

\begin{figure}[htbp]
    \includegraphics[width=0.48\textwidth]{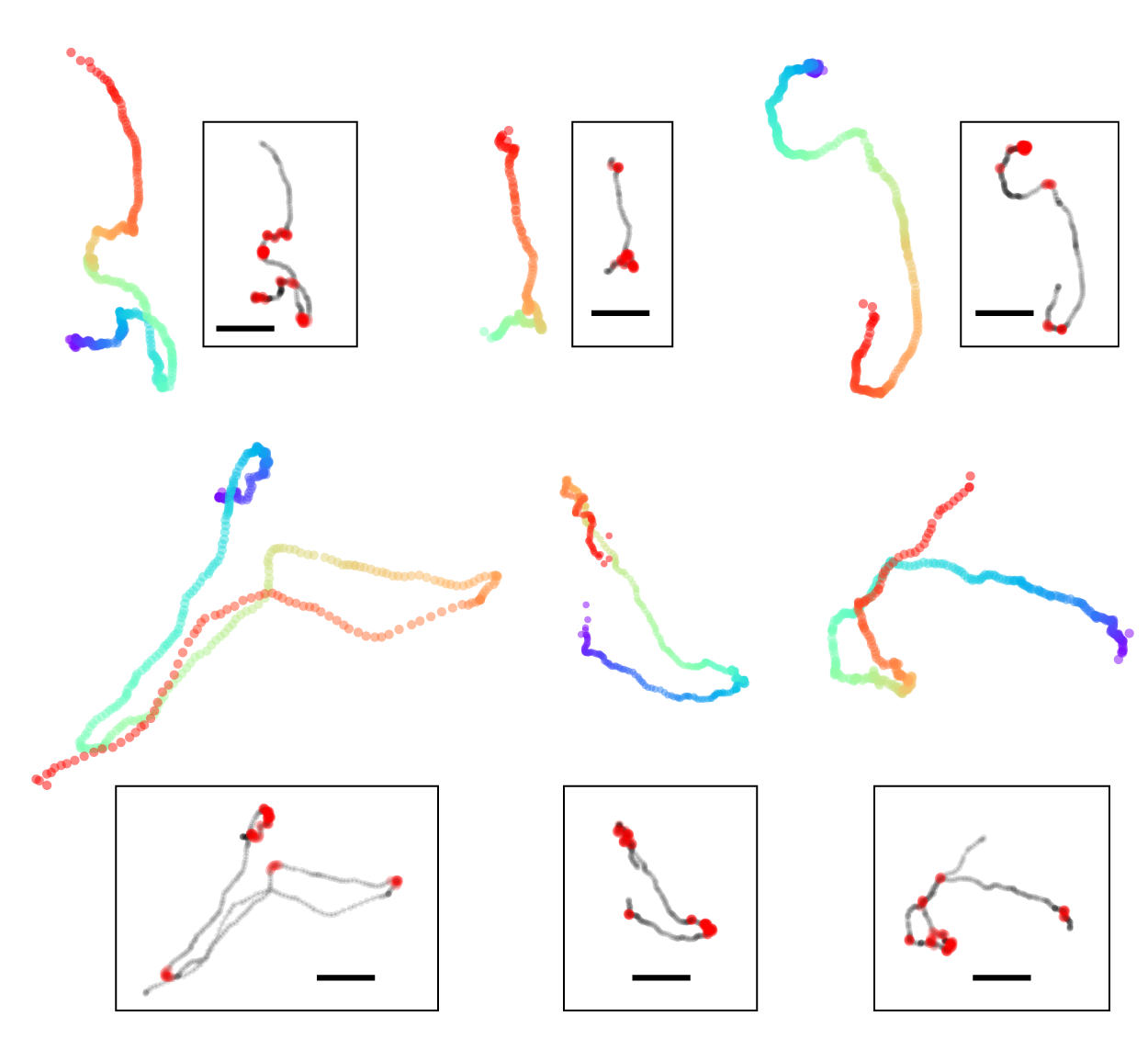}
    \caption{Typical tracks of MCF-10A cells examined on collagen. The colors (from red to blue) represent the time. Turns are marked in insets. All the displayed tracks contains at least one long turn event ($\geq$30 min). Scale bars in the insets: 20 \um.} \label{fig:demo_track}
  \vspace{-3mm}
\end{figure}

Finally, we display some typical tracks with turns marked, see Fig.~\ref{fig:demo_track}. The shown tracks all contain at least one long turn event ($\geq30$ min). Turns shorter than 6 min (3 frames) are not displayed in the boxes. With these shown tracks, we caution readers against the assumption that cells turning at a constant rate produce arc-like or circular corners in their trajectories~\cite{Abaurrea2017,Allen2020}. This scenario requires a stable speed of propagation during turns but it is not the case for the cells examined here.

\section{Angular changes during runs}

\begin{figure}
    \centering
    \includegraphics[width=0.55\linewidth]{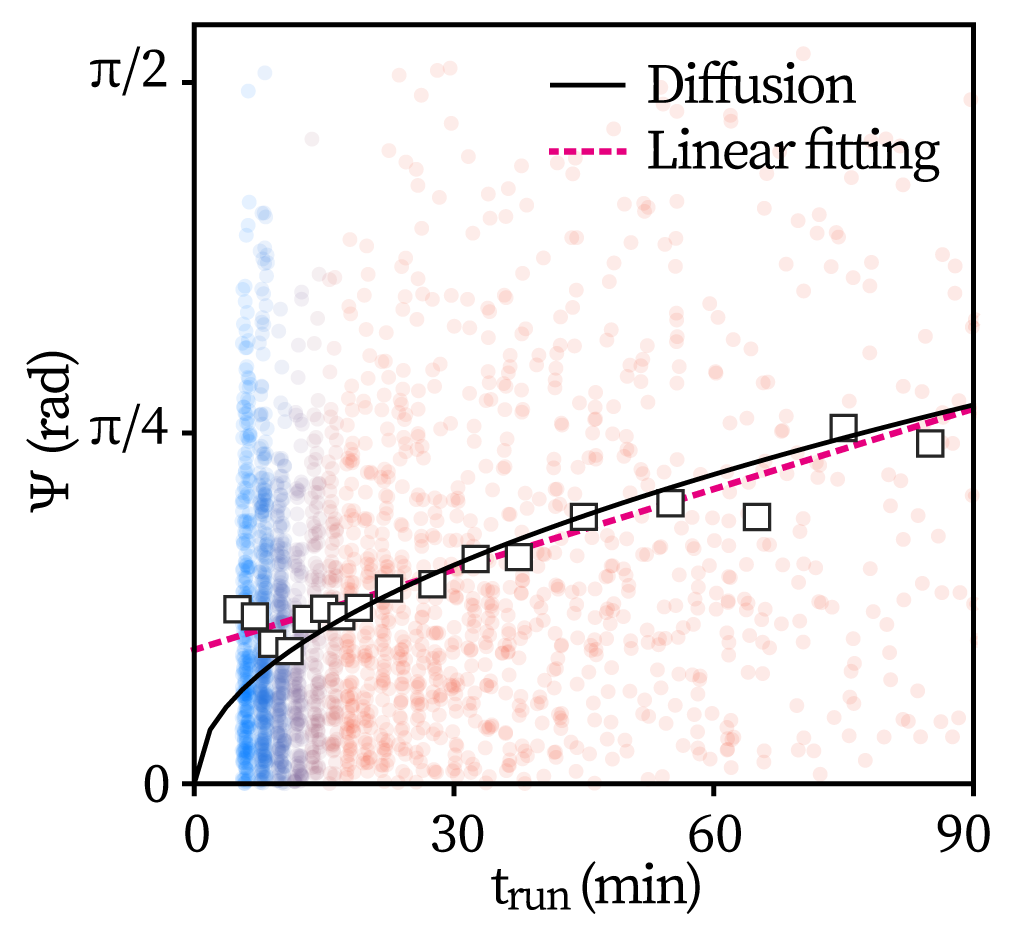}
    \caption{Zoom-in of the data presented in Fig.~\ref{fig:rotor}(d) in the main text, with diffusive fitting.}
    \label{fig:diffusiveRun}
\end{figure}

During runs, angular changes accumulate at slow rates. Estimating from the linear trends shown in Figs.~\ref{fig:rotor}(d), \ref{fig:mesenchymal}(d), and \ref{fig:mesenchymal}(i), the turning rates read 0.006, 0.007, and 0.005 rad/min, respectively. This turning rate is comparable to angular diffusion induced by environmental noise. In Fig.~\ref{fig:diffusiveRun}, we present the same data as in Fig.~\ref{fig:rotor}(d) and fit $\langle \Psi \rangle$ with $\sqrt{2D_{r}t}$. We obtain $D_{r,\rm run}=0.004$~min$^{-1}$. The result aligns well with $D_{r,\rm run}=0.004-0.005$~min$^{-1}$ obtained from fitting ACFs [Figs.~\ref{fig:rotor}(e), \ref{fig:mesenchymal}(e), and \ref{fig:mesenchymal}(j)], and is equivalent to the thermal diffusivity of a cell-sized sphere in water $D_r=k_BT/8\pi\mu R^3\approx0.004$~min$^{-1}$. $k_B$ is the Boltzmann constant, $T=310$ K is the experimental temperature, $\mu=0.69$~mPa$\cdot$s is the viscosity of water under this temperature, and $R=15$~\um is the cell radius. These results suggest that angular changes during runs may derive from thermal diffusion.

\section{Turning sign and rate on the single-cell level}

\begin{figure}[htbp]
    \includegraphics[width=0.48\textwidth]{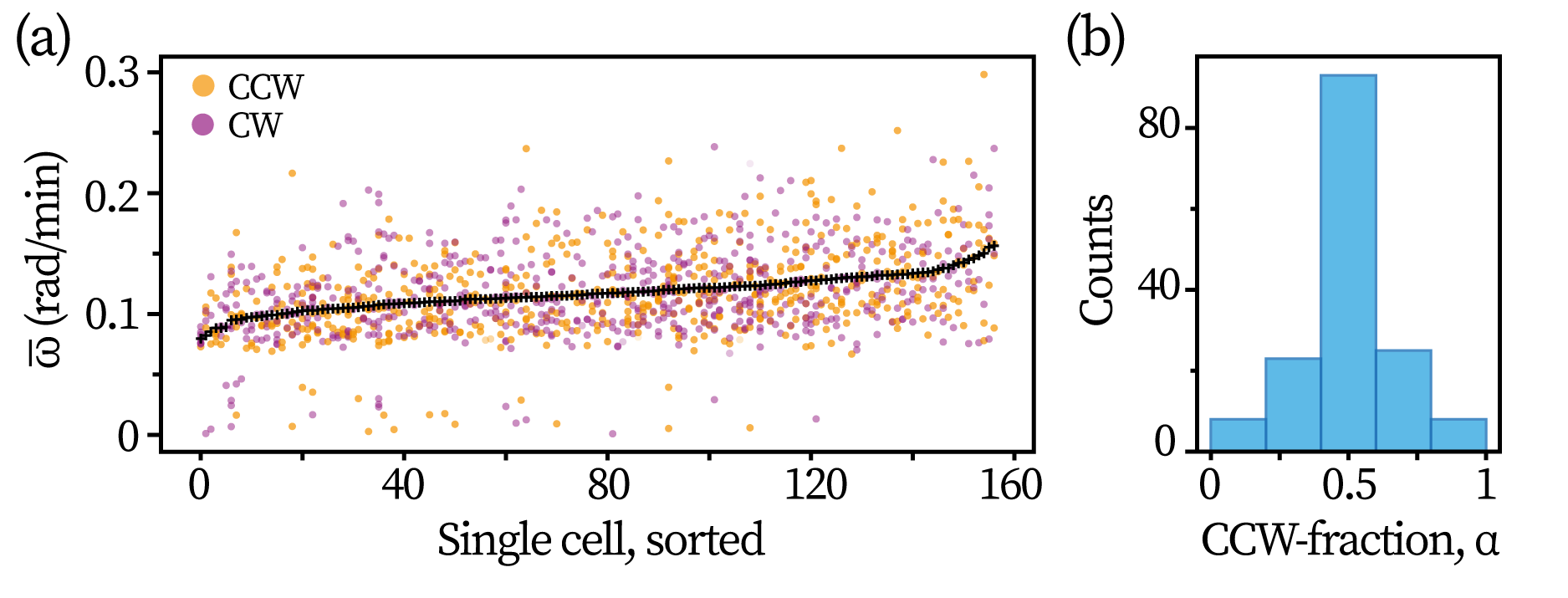}
    \caption{Turning rate and chirality on the ensemble and individual level for MCF-10A cells on collagen. (a) Cells ($N$=157) that display more than 5 turns in the recording are sorted and displayed along the $x$-axis, from left to right, by their respective mean turning rate (black line). Dots represent single turns and are colored by the chirality. (b) Frequency counts of single-cell CCW-fraction, see text for detailed definition.} \label{fig:chirality}
  \vspace{-3mm}
\end{figure}

So far, all turns are pooled on an ensemble level, without information of the how each individual cell perform turns.
It is especially of interest to know, if on the single-cell level, a cell would still turn on a near-constant rate for its consecutive turns. For this, we collect cells ($N$=157) that display multiple ($\geq$5) turns and present the mean absolute turning rate of each turn $\overline{\omega} = \int_0^{t_{\rm turn}}\omega dt/t_{\rm turn}$, alongside with the turns' chirality, see Fig.~\ref{fig:chirality}(a). For the clarity of display, the cells are sorted by their mean turning rate over the multiple turns. It is obvious that, the turning rate for a single cell is almost as scattered as the turning rates on the ensemble level. The microscopic picture is more likely that the absolute turning rate of a turn for all cells are following nearly the same distribution. 

In Fig.~\ref{fig:taxis}(e) we have shown that, there is no preference of the sign of turning (chirality) on the ensemble level. We resolve how does this neutrality hold for individual cells. For each cell, we calculate the fraction of counter-clockwise (CCW) turns $\alpha = N_{\rm CCW}/(N_{\rm CCW}+N_{\rm CW})$ with $N_{\rm CW}$ the number of clockwise turns. Figure~\ref{fig:taxis}(a) shows the distribution of $\alpha$ over the cells. The overall distribution is symmetric around ($\alpha=0.5$) with
A dominant portion ($>50\%$) of cells displaying both turning signs equally likely (the central bin).  

\section{Persistence from a runner-roter perspective}
Here we detail the implications of runner-rotor kinematics of adherent cells for  conventional practices in measuring persistence.

\begin{figure}[htbp]
    \includegraphics[width=0.48\textwidth]{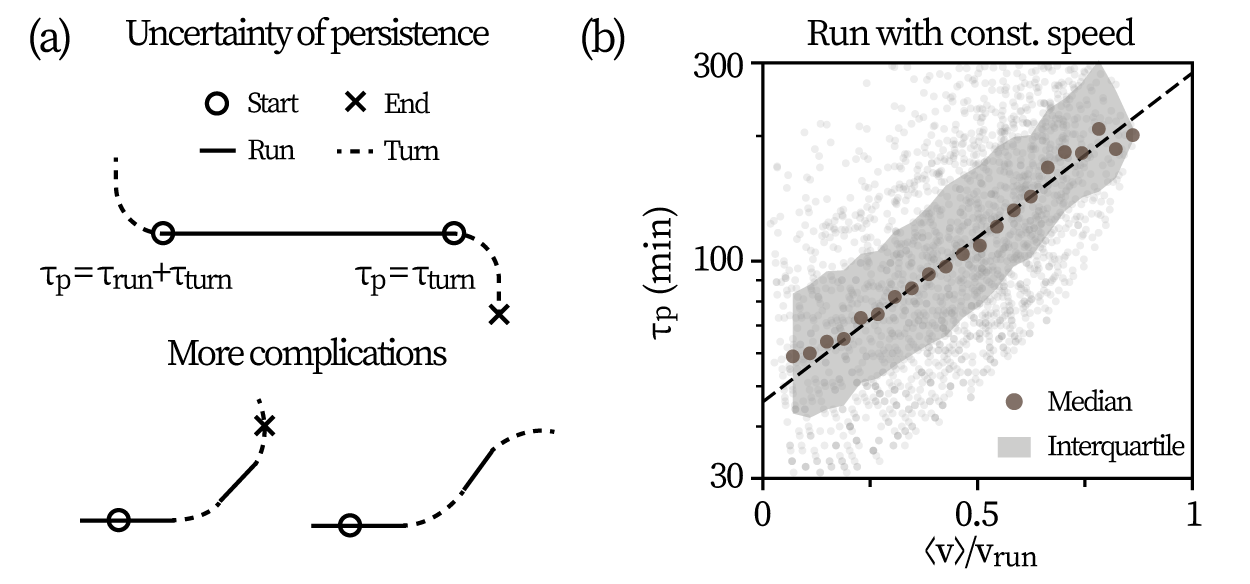}
    \caption{Conventional practice in measuring persistence and demonstrating the speed-persistence can be problematic. (a) Depending on the starting point of measurement, the reported persistence may vary greatly. (b) Persistence defined in conventional way measured in simulated tracks, where the speed is simply a constant and not coupled to persistence. Parameters used: $\tau_{\rm run}=\tau_{\rm turn}=25$ min, $dt=1$ min, $\Omega_{\rm turn}=\pi/2\tau_{\rm turn}$, $v_{\rm run}=1$ \umm, and $v_{\rm turn}=0.05$ \umm. The motion is free of noise.}\label{fig:challenge}
  \vspace{-3mm}
\end{figure}

Consider a simplified case where a cell runs for $\tau_{\rm run}$ at the same direction, and then turns 90\degree~in the coming time $\tau_{\rm turn}$. Then if measurement starts at the beginning of a run, the reported persistence $\tau_p=\tau_{\rm turn}+\tau_{\rm run}$; or if measurement starts at the end of a run, $\tau_p=\tau_{\rm turn}$, see Fig.~\ref{fig:challenge}(a) upper panel. One should note the high uncertainty in this measurement as the upper extreme $\tau_{\rm turn}+\tau_{\rm run}$ is typically several times larger than the lower extreme $\tau_{\rm turn}$. More complications may be induced when the cell turns, on average, for less than the threshold (90\degree) each time, see the lower panels of Fig.~\ref{fig:challenge}(a). In this case, $\tau_p$ is likely to include more than 1 times of $\tau_{\rm turn}$. In fact, when two consecutive turns can be of opposite directions, $\tau_p$ may include infinite times of $\tau_{\rm turn}$.

The second challenge brought up by the runner-rotor model is the conventional way of demonstrating UCSP with an exponential dependence between persistence time and speed~\cite{Maiuri2015}. In simulated runner-rotor tracks, we find such dependence to manifest even without any coupling between persistence and speed, Fig.~\ref{fig:challenge}. The simulated runner-rotor particle moves at a constant speed $v_{\rm run}$ along a given direction in runs; and turns at a constant rate $\Omega_{\rm turn}$ with no displacement during turns. We neglect all stochastic elements, such as fluctuations in velocity, angular velocity, and movement direction, except for the random transitions between the particle's running and turning states. The persistence $\tau_p$, following the conventional definition, is measured from randomly chosen starting points. Results show an exponential dependence between the reported persistence and the mean instantaneous speed $\langle v \rangle$, see the dashed line in Fig.~\ref{fig:challenge}. By modulating the time fraction of runs and turns ($\tau_{\rm run}/\tau_{\rm turn}$), simulation results could exhibit different trends, many of which feature an exponential dependence between $\tau_p$ and $\langle v \rangle$. 

This minimal simulation scheme is as follows.
\begin{enumerate}
    \item The particle's state at the $i$-th time step is described by the vector $(s_i, c_i, \bm{x}_i, \phi_i)$, where $s_i = \{\rm{run, turn}\}$ represents the motility state, $c_i = \pm 1$ represents the chirality, $\bm{x}_i$ is the particle's location vector, and $\phi_i$ is its direction.

    \item At the $i$-th time step, $s_i$ is first determined. The probability for the previous state $s_{i-1}$ to continue is $p = e^{-dt / \tau_{s_{i-1}}}$.

    \item Only when $s_{i-1} = \rm{run}$ and $s_{i} = \rm{turn}$, $c_i$ is randomly assigned a value of $+1$ or $-1$. For other scenarios, $c_i = c_{i-1}$.

    \item Next, the particle's direction is updated:
    \[
    \phi_i = 
    \begin{cases}
        \phi_{i-1} & \text{if } s_i = \rm{run}, \\
        \phi_{i-1} + c_i \Omega_{\rm turn} dt & \text{if } s_i = \rm{turn}.
    \end{cases}
    \]

    \item Then, the particle moves according to its updated direction:
    \[
    \bm{x}_i = \bm{x}_{i-1} + v_{s_i} dt \cdot \bm{n},
    \]
    where $\bm{n} = (\cos\phi_i, \sin\phi_i)$ is the unit direction vector.

    \item Proceed to the next time step.
\end{enumerate}

\bibliographystyle{naturemag}
\bibliography{./reference.bib}

\end{document}